\documentclass[fleqn,10pt]{wlscirep}
\usepackage[utf8]{inputenc}
\usepackage[T1]{fontenc}
\usepackage{graphicx}
\usepackage{mwe}
\usepackage{subfigure}
\usepackage{url}

\title{Understanding the complex dynamics of climate change in south-west Australia using Machine Learning} 
\author[1,]{Alka Yadav}
\author[2,*]{Sourish Das}
\author[3]{K Shuvo Bakar}
\author[1,4,*]{Anirban Chakraborti}

\affil[1]{Jawaharlal Nehru University, School of Computational and Integrative Sciences, New Delhi-110067, India}
\affil[2]{Chennai Mathematical Institute, Chennai-603103, India}
\affil[3]{School of Public Health, The University of Sydney, NSW 2006, Australia}
\affil[4]{BML Munjal University, School of Engineering and Technology, Gurugram-122413, India }
\affil[*]{sourish@cmi.ac.in}
\affil[*]{anirban.chakraborti@bmu.edu.in }


\begin{abstract}
The Standardized Precipitation Index (SPI) is used to indicate the meteorological drought situation -- a negative (or positive) value of SPI would imply a dry (or wet) condition in a region over a time period. The climate system is an excellent example of a ``complex system'' since there is an interplay and inter-relation of several climate variables. It is not always easy to identify the factors that may influence the SPI, or their inter-relations (including feedback loops). Here, we aim to study the complex dynamics that SPI has with the sea surface temperature (SST), El Ni\~{n}o Southern  Oscillation (ENSO) (aka., NINO 3.4) and Indian Ocean Dipole (IOD), using a machine learning approach. Our findings are: (i) IOD was negatively correlated to SPI till 2008; (ii) until 2004, SST was negatively correlated with SPI; (iii) from 2005 to 2014, the SST had swung between negative and positive correlations; (iv) since 2014, we observed that the regression coefficient ($\delta$) corresponding to SST has always been positive; (v) the SST has an upward trend, and the positive upward trend of $\delta$ implied that SPI has been positively correlated with SST in recent years; and finally, (vi) the current value of SPI has a significant positive correlation with a past SPI value with a periodicity of about 7.5 years. Examining the complex dynamics, we used a statistical machine learning approach to construct an inferential network of these climate variables, which revealed that SST and NINO 3.4 directly couples with SPI, whereas IOD indirectly couples with SPI through SST and NINO 3.4. The system also indicated that Nino 3.4  has a significant negative effect on SPI. Interestingly, there seems to be a structural change in the complex dynamics of the four climate variables of NINO 3.4, IOD, SST, and SPI, some time in 2008. 
Though a simple 12-month moving average of SPI has a negative trend towards drought, the complex dynamics of SPI with other climate variables indicate a wet season for western Australia.
\end{abstract}

\begin{document}
\flushbottom
\maketitle

\section*{Introduction}

Global climate change affects the large-scale atmospheric circulation anomalies, resulting in metrological drought in many parts of the world, including Australia \cite{METEOROLOGICALDROUGHT,Heathcote1969}. Agricultural system is highly vulnerable with the extreme volatility of climate variables. In particular, in Western Australia the wheat \cite{Ludwig2009} and broadacre livestock \cite{waclimate} productions are in concern. The agriculture water supplies have decreased quite drastically, which is about 44\% reduction in 2010-2018 compared to 2001-2009 \cite{waclimatetrends}. According to the Bureau of Meteorology (BoM), the annual mean temperature anomaly is increased with a fluctuation in the early and late years of the millennium \cite{waclimatechange}. Similar pattern has also been observed for the rainfall anomaly during this time \cite{slowingtrend}. Understanding the complex dynamics of the drought with respect to climate change and its impacts on the ecosystem is essential, as it severely impacts agriculture, water resources and public health \cite{VICENTESERRAN,Sugg,Huai2017,Wang2019}. In this paper, we study the complex dynamics between climate variables, such as sea surface temperature (SST), El Ni\~{n}o Southern Oscillation (ENSO) NINO 3.4, Indian Ocean Dipole (IOD) and their impact on the standard precipitation index (SPI) in south-west Australia using the Machine Learning (ML) based approach. Western Australia contributes 18\% (about 10.7 billion AUD) of the total gross agricultural production in Australia\cite{footnotet1}, and the grain exports are worth around 4 billion AUD\cite{footnote2}. There are four types of droughts: meteorological, hydrological, agricultural, and socioeconomic \cite{Heim}. Among them, meteorological droughts are characterized by below-normal precipitation and measured by SPI. It explains the degree of dryness and the duration of the dry period. Usually, low SPI is prone to trigger other types of droughts.

To understand meteorological drought, McKee \emph{et al.} (1993) developed SPI using rainfall measurements \cite{McKee1993,Christos2011,KEMAL2005}. Wenhong \emph{ et.al} (2008) showed that the SPI over the southern Amazon region decreased from 1970 through 1999 by 0.32 per decade, indicating an increase in dry conditions \cite{Wenhong2008}. Rengung \emph{et. al.} (2008) presented their study based on the SPI and analyzed the relationship of summer droughts in the United States \cite{Rengung2008}. Climatic factors also influence meteorological droughts, such as a brief review that can be found on the  ENSO variability and drought risk over Australia \cite{Francois2020}. The dynamics of ENSO and IOD cycles with the pattern of occurrence of meteorological drought is reported \cite{inbook}.
The direct impact of ENSO on precipitation was reported by Wen \emph{et al.} (2015) \cite{Wen2105}. 

Loughran \emph{et. al.} (2018) investigated how the ENSO couple the mechanisms of heatwaves in Australia  \cite{Loughran2018}. They examine the ENSO large-scale mode of variability that influences the Australian heatwaves using prescribed SST characteristics of El Ni\~{n}o and La Ni\~{n} conditions\cite{Loughran2018}. Lee \emph{et al.} (2009) showed that when the time lag is 0 or 1 month, the November-February ENSO, SST explains much of the drought signals over eastern Australia \cite{Lee2009}. Taschetto \emph{et al.} (2009) investigated the inter-seasonal impact of ENSO on Australian rainfall using peak SST anomalies \cite{Taschetto2009}, and the correlation between ENSO and the eastern part of the IOD is positive from January to June, and then changes to negative from July to December \cite{Koll2011}.  Takeshi \emph{et al.} (2014) showed that the IOD could affect the ENSO state, in addition to the well-known preconditioning by equatorial Pacific warm water volume \cite{zumo2014}. It also explored the interdecadal robustness of this result over the 1872 to 2008  \cite{zumo2014}. SST measurement is an essential factor that influences the Australian rainfall pattern \cite{Nicholls1989}. Nicholls \emph{et. al.} (1989) presented an ML-based approach using the rotated principal component analysis of Australian winter (June-August) rainfall, which revealed the correlation of precipitation and SST in the Indian and Pacific oceans \cite{Nicholls1989}. Similarly, other studies analyzed the link between SST, SPI, and global vegetation \cite{Sietse2001,Wasyl2001,Lee2009,Heathcote1969}.

There are sparse multivariate approaches to understanding the complex dynamics of drought available in the literature. Most of the studies focused on the pairwise relationship between the NINO 3.4 with SST, or NINO 3.4 with IOD separately, or their pairwise effect on SPI \cite{Loughran2018,Lee2009}. In this paper, we explored the combined teleconnection of SST, NINO 3.4 and IOD on SPI through ML based approach. Here, we developed the Granger causal model to see the causal behaviour of SST, IOD and NINO 3.4 with SPI and then their effect on SPI altogether in south-west Australia.

\section*{Data}\label{sec:data}

\begin{figure}[p]
\centering
\begin{tabular}{cc}
    \includegraphics[width=.4\linewidth]{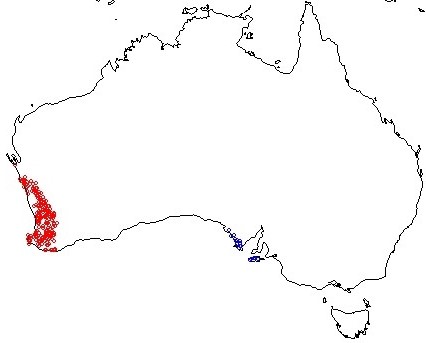}
 &  \includegraphics[width=.5\linewidth]{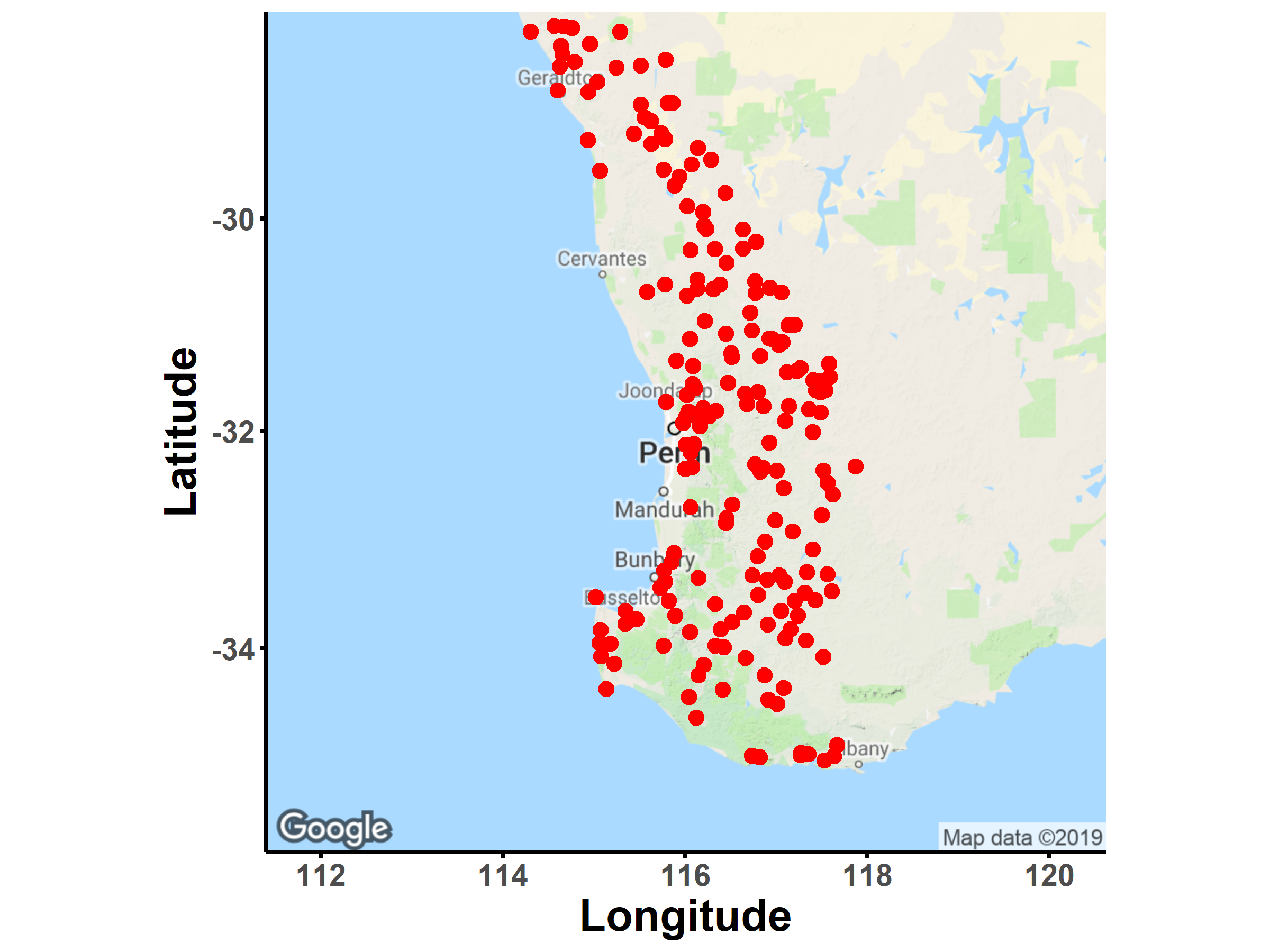}\\
 (a) & (b)\\
\end{tabular}
\caption{(a) Map of Australia with the rainfall gauged locations is superimposed in the study area. (b) A close representation of the 194 gauged locations in the south-west part of Australia. The range of the area considered varies from longitude position 113.72 to 137.12 and latitude position -26.70 to -35.73. }
\label{fig_lacation_on_Australian_map}
\end{figure}

We used four variables for this study: SPI,  SST,  NINO 3.4 and IOD indices for the Western Australia region. The NINO 3.4 and IOD  data are available from NOAA website\cite{footnotet3}.
The monthly SPI is calculated from the daily precipitations observed in 194 stations in south-west Australia, which is downloaded from the Bureau of Meteorology (BOM) website\cite{footnotet4}. We have extracted the SST data for the Australian region from NOAA \cite{SST_Data_Source}. The range of the area we have considered for this study varies from longitude position 113.72 to 137.12 and latitude position -26.70 to -35.73. Figure~(\ref{fig_lacation_on_Australian_map}) presents the 194 rainfall gauged locations on the map of Australia. For this study, we have used SPI monthly time series over 58 years (1961-2018). Since SST data is available from 1982, we used SST, IOD and Nino 3.4  monthly averaged data from 1982 to 2018 and performed all our model analyses for this period. 

\subsection*{SPI for south-west Australia}

The SPI is a popular index to monitor the drought. In 1993, McKee \emph{et al.} suggested a classification scale for SPI \cite{McKee1993,Hayes1999}. For any given location, SPI values are classified into seven different regimes (from wet to dry), as shown in Table (\ref{tab_SPI_range}). The SPI is a unitless index in which negative values indicate drought, where $-1$ is commonly used as a threshold; positive values mean wet conditions.

\begin{table}[p]
\caption{Classification scale for Standardized Precipitation Index (SPI).}
	\label{tab_SPI_range}
	\centering
\setlength{\arrayrulewidth}{1mm}
\setlength{\tabcolsep}{18pt}
\renewcommand{\arraystretch}{1.5}
\begin{tabular}{p{2.5cm}p{2.5cm}}\hline
  \bf{SPI range} &\bf{ Category} \\	\hline
2.00 and above & Extremely wet \\
1.50 to 1.99 & Very wet \\
1.00 to 1.49 & Moderately wet\\
- 0.99 to 0.99& Near normal\\
- 1.00 to -1.49&Moderately dry\\
- 1.50 to -1 . 99&Severely dry\\
-2.00 and less&Extremely dry\\  \hline
\end{tabular}
\end{table}

\begin{figure}[p]
    \centering
    \includegraphics[width=0.6\linewidth]{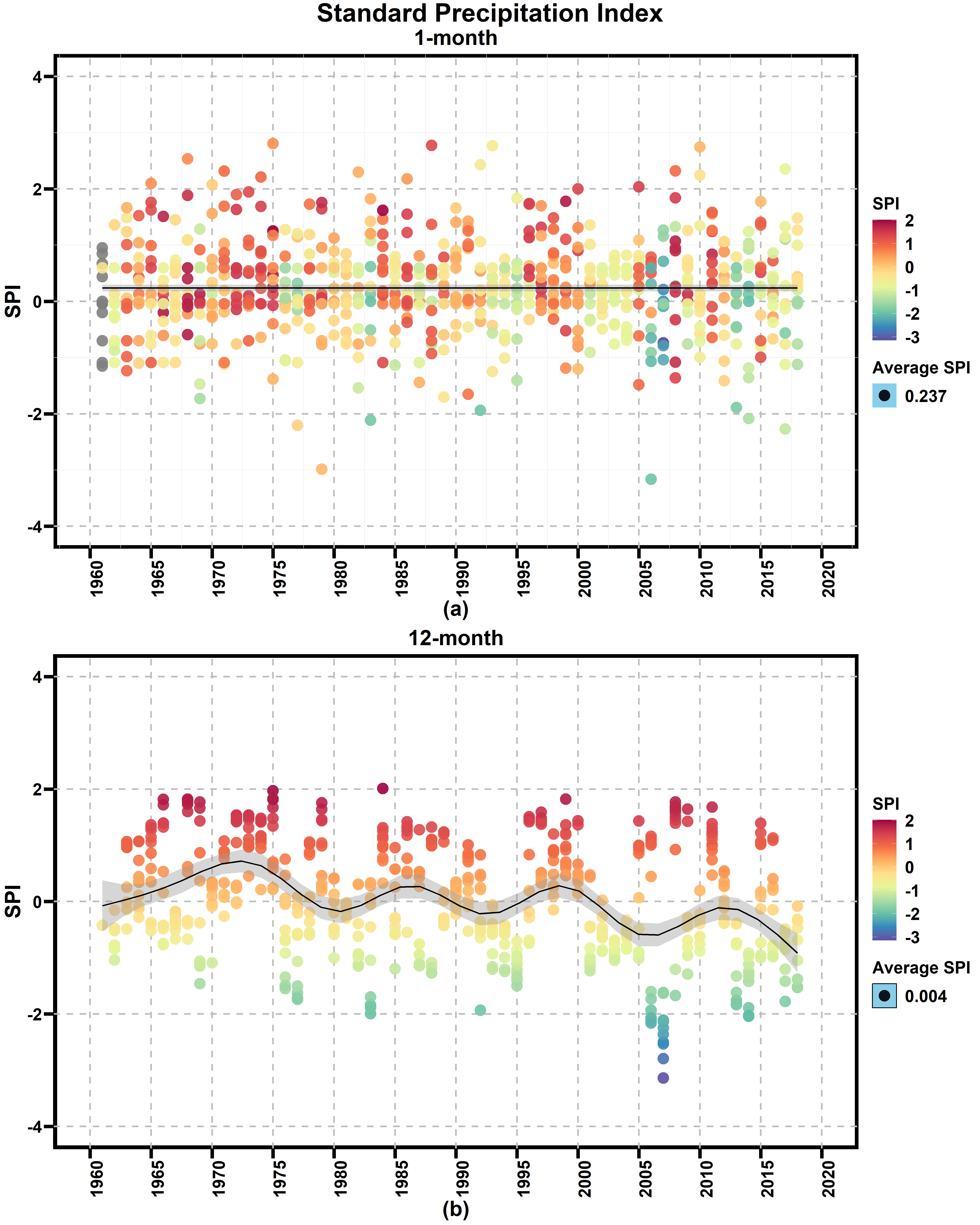}
     \caption{The Standardized Precipitation Index (SPI) (a) one-month and (b) 12-month time series over 58 years (1961-2018) period with the 12-month moving average (solid black line). A high signal-to-noise ratio helps reduce the standard error of our analysis \cite{Lehner2006}. We see that the signal-to-noise ratio is much higher for SPI 12-month series than for the SPI 1-month series. We checked this for all 194 monitoring stations and saw a similar higher signal-to-noise ratio for the 12-month series. Hence,  we considered the SPI 12-month series for our analysis. }
     \label{fig_ts_SPI}
\end{figure}

Figure (\ref{fig_ts_SPI}a) presents the SPI 1-month time series over 58 years (from 1961 to 2018) for 194 locations in south-west Australia. Figure (\ref{fig_ts_SPI}b) presents the monthly SPI calculated with a 12-month epoch and annual moving average for the site positioned at longitude $113.72$ and latitude $-26.70$. From the plots, we see that the signal-to-noise ratio is much higher for SPI 12-month series than for the SPI 1-month series. We checked this for all 194 monitoring stations and saw a similar higher signal-to-noise ratio for the 12-month series. Hence, the rest of the analysis is based on considering the SPI 12-month series. A high signal-to-noise ratio helps reduce the standard error of our analysis \cite{Lehner2006}.

\begin{figure}[p]
\centering
\begin{tabular}{c|c}
\includegraphics[width=.45\textwidth]{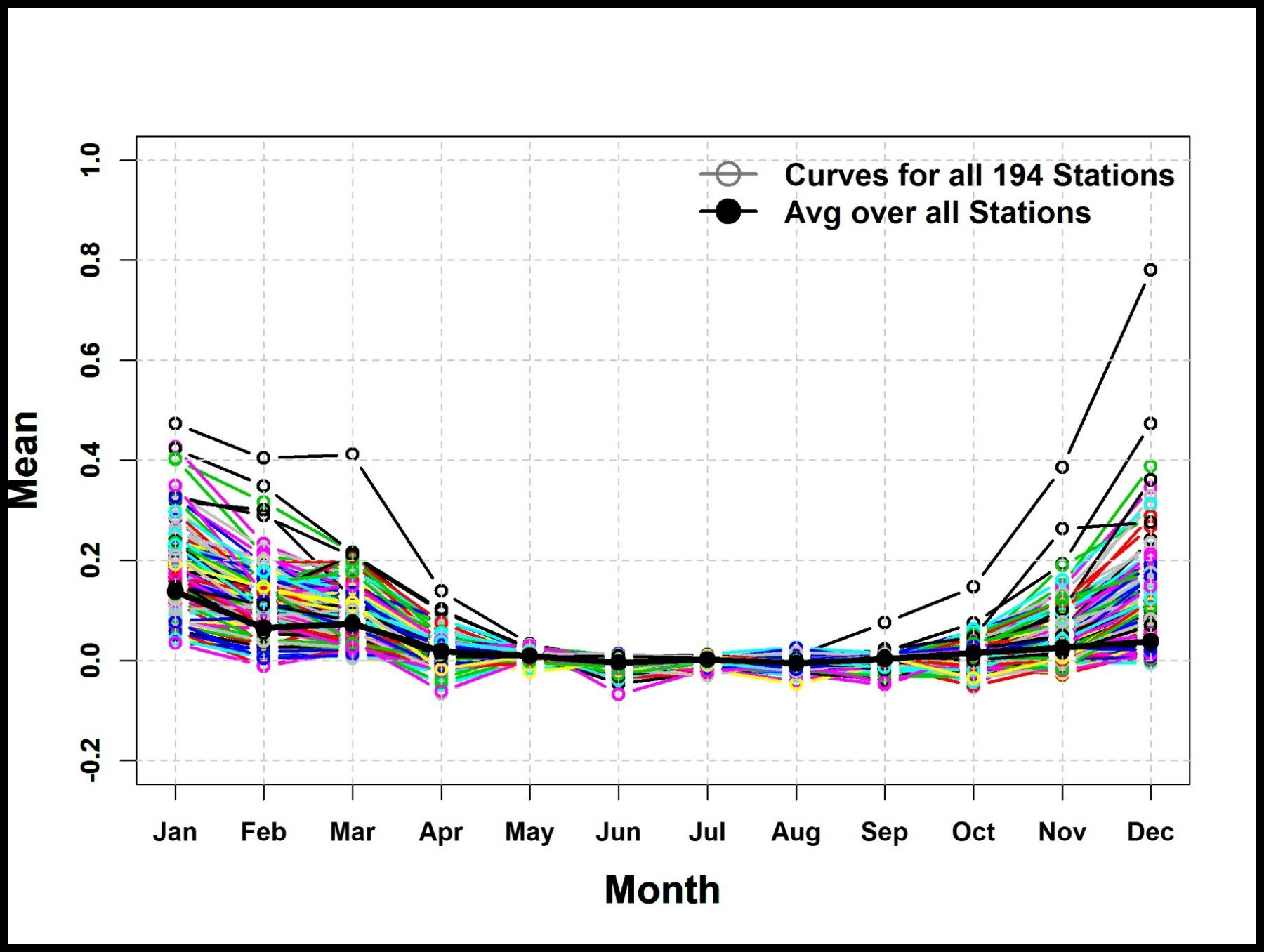}
& \includegraphics[width=.5\textwidth]{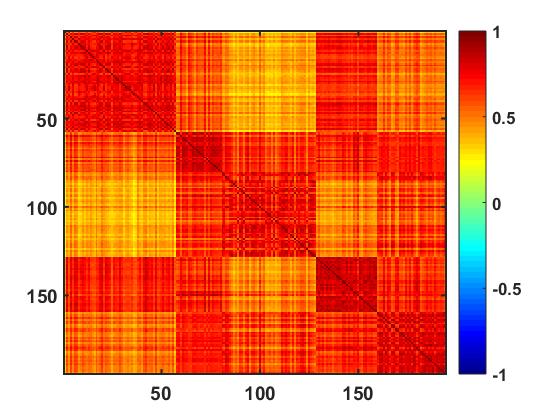}\\
(a) & (b)\\
\end{tabular}
\caption{(a) Plot of the average value of Standardized Precipitation Index (SPI) for all months over 58 years (from 1961 to 2018) in each location in south-west Australia. The dark black curve is the average of all 194 curves. (b) Visual representation of the correlation matrix of SPI among 194 locations. This correlation matrix depicts that all the locations are spatially correlated.}
\label{fig_avg_by_month}
\end{figure}

In south-west Australia, October to March is monsoon time. Hence we observe higher variability of SPI compared to other months. Figure (\ref{fig_avg_by_month}a) shows the average SPI for each month over 58 years (from 1961 to 2018) for all 194 sites. Figure (\ref{fig_avg_by_month}b) is a correlation matrix among all 194 stations over all 58 years. Visual representation of the correlation matrix of SPI depicts that all the locations are spatially correlated.

\section*{Results}\label{sec:results}
 
\begin{figure}[p]
    \centering
    \includegraphics[width=.75\linewidth]{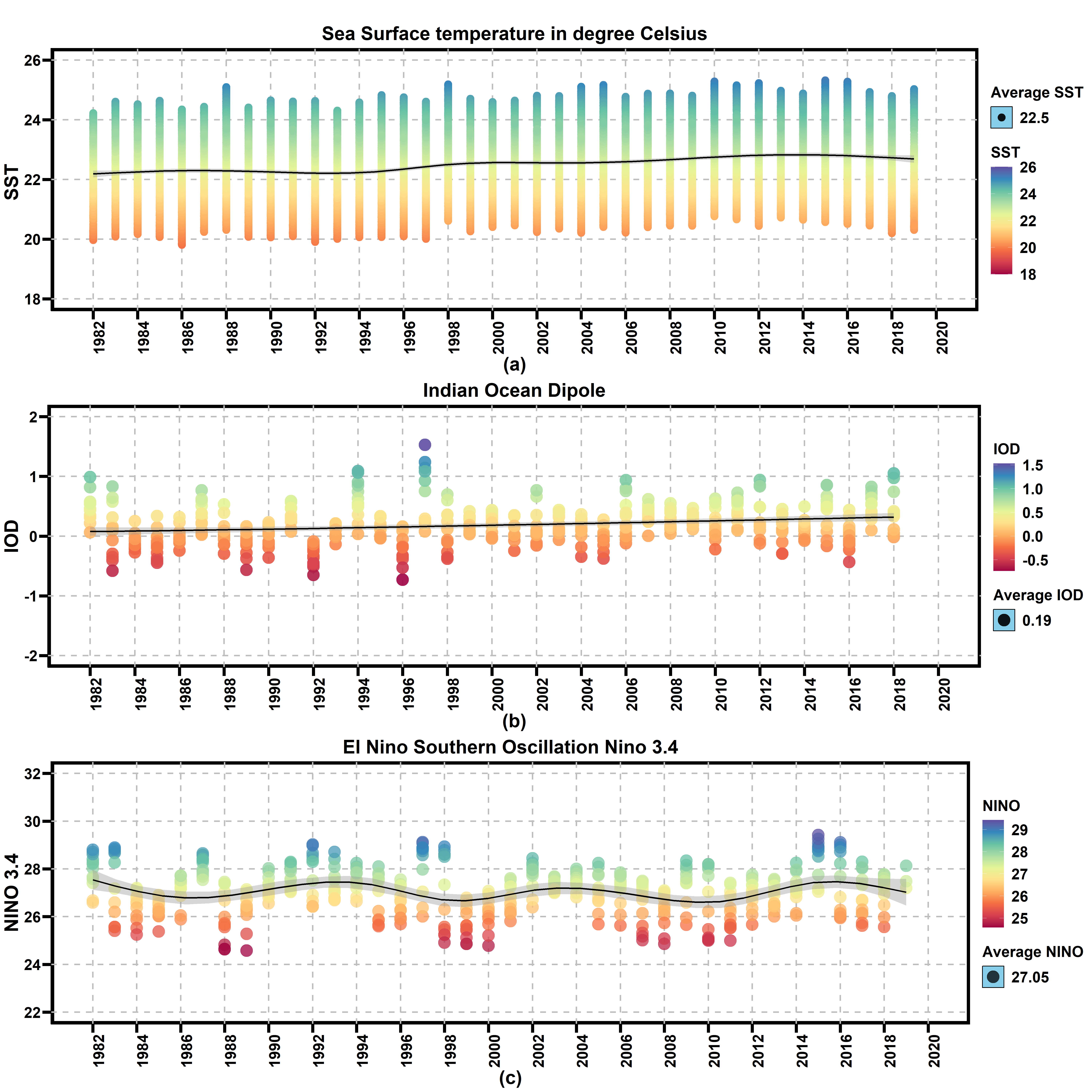}
    \caption{Time-series plots for (a) Sea Surface Temperature (SST) (b) Indian Ocean Dipole (IOD) and (c) El Niño–Southern Oscillation (ENSO) NINO 3.4 over 37 years (1982-2018) for south-west Australia. The black curve is the three years moving average.}
    \label{ts_nino_sst_iod}
\end{figure}

We explored the dynamics of SPI with SST, NINO 3.4 and IOD over 1982 to 2018 period. Figures (\ref{ts_nino_sst_iod}) present the time series of SST, NINO 3.4, and IOD respectively. We see that these are mean reversal processes, i.e. the series values return to their means after a certain period. To check whether these series have a long memory or not, we estimated the Hurst exponent, which relates to the autocorrelations of the time series. The Hurst exponent coefficient\cite{Hurst2011} in Table (\ref{tab_Hurst_exp}) explains that the variables, including the SPI, have long memory as the values are above 0.5.

\begin{table}[p]
\caption{This table presents the Hurst exponent value for all index, i.e., SPI, IOD, SST and Nino 3.4. The values indicate that the system has a long memory.}
\label{tab_Hurst_exp}
	\centering  
\setlength{\arrayrulewidth}{1mm}
\setlength{\tabcolsep}{18pt}
\renewcommand{\arraystretch}{1.5}
\begin{tabular}{p{7cm}p{3cm}} \hline
	\bf{Index}&\bf{Hurst Value} \\
	\hline
	Standard Precipitation Index (SPI)  &  0.71 \\
	El Ni\~{n}o Southern Oscillation (ENSO) NINO 3.4  & 0.66\\
	Indian Ocean Dipole (IOD)  &   0.69    \\
	Sea Surface Temperature (SST)& 0.58 \\ \hline
	\end{tabular}
\end{table}

\begin{figure}[p]
    \centering
    \includegraphics[width=0.75\linewidth]{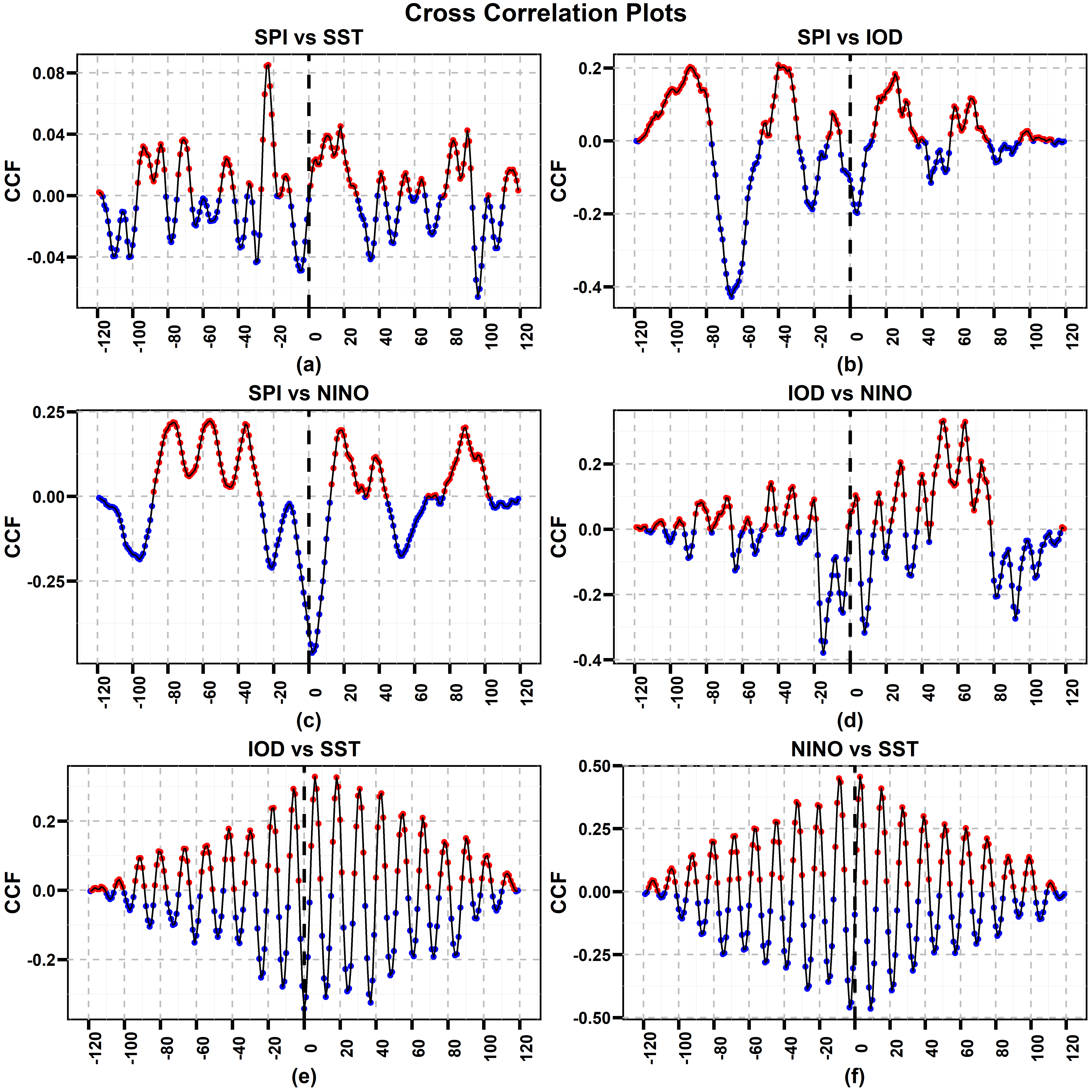}
    \caption{For decade: 1999 to 2008; The cross-correlation function (CCF) plots between (a) Standard Precipitation Index (SPI) and Sea Surface Temperature (SST), (b) SPI and Indian Ocean Dipole (IOD), (c) SPI and Nino 3.4, (d) IOD and Nino 3.4, (e) IOD and SST, (f) Nino and SST with a lag of 120. The  cross-correlation coefficient is significant enough to tell that all these indices have significant effect on each other.}
    \label{fig_ccf_1999_2008}
\end{figure}

\begin{figure}[p]
    \centering
    \includegraphics[width=0.75\linewidth]{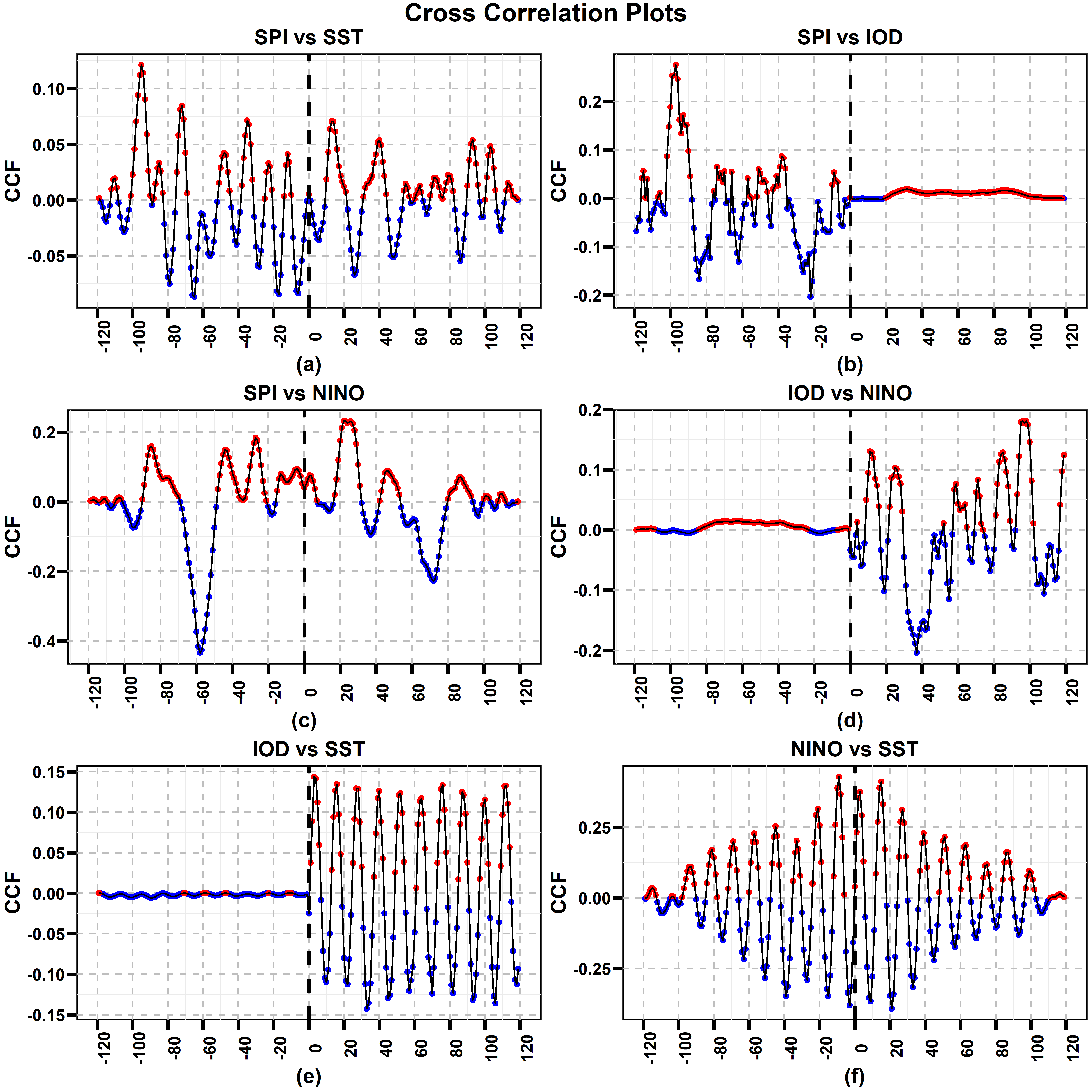}
    \caption{For decade: 2009 to 2018; The cross-correlation function (CCF) plots between (a) Standard Precipitation Index (SPI) and Sea Surface Temperature (SST), (b) SPI and Indian Ocean Dipole (IOD), (c) SPI and Nino 3.4, (d) IOD and Nino 3.4, (e) IOD and SST, (f) Nino and SST with a lag of 400. It shows that though the ccf value is small but SPI and SST are correlated to each other. SPI has an effect on IOD but vice versa is not true. SPI and IOD are not that significantly correlated to each other. It is clear that both SPI and NINO 3.4 are significantly correlated to each other. From the figure it is clear that IOD has an effect on NINO 3.4 but reverse is not true. IOD has a small effect on SST but SST has no effect on IOD. We see that both NINO 3.4 and SST are significantly correlated to each other. The cross-correlation coefficient is significant enough to tell that all these indices have significant effect on each other.}
    \label{fig_ccf_2009_2018}
\end{figure}

\begin{figure}[p]
\centering
\begin{tabular}{cc}
    \includegraphics[width=.5\linewidth]{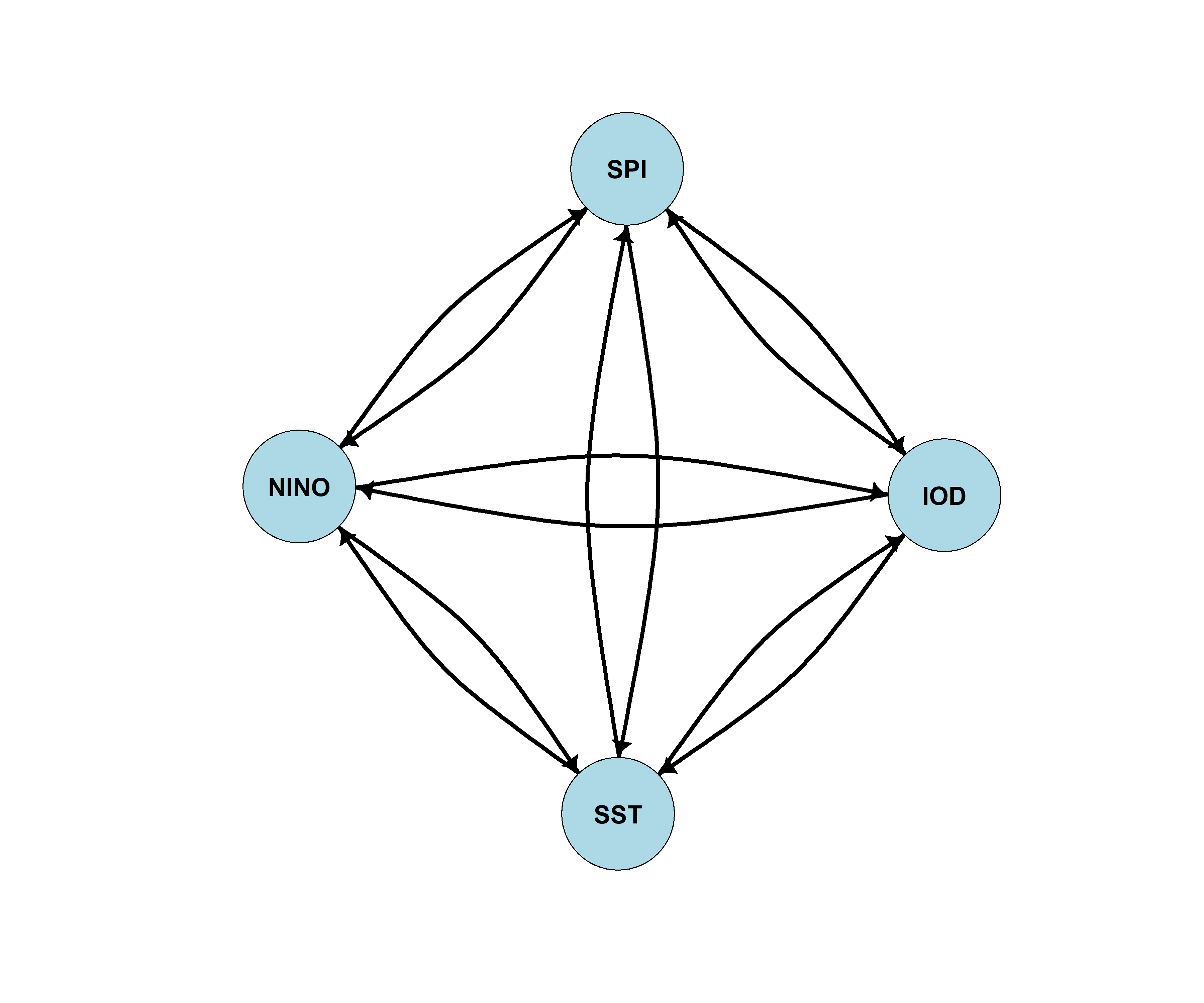}
 &  \includegraphics[width=.5\linewidth]{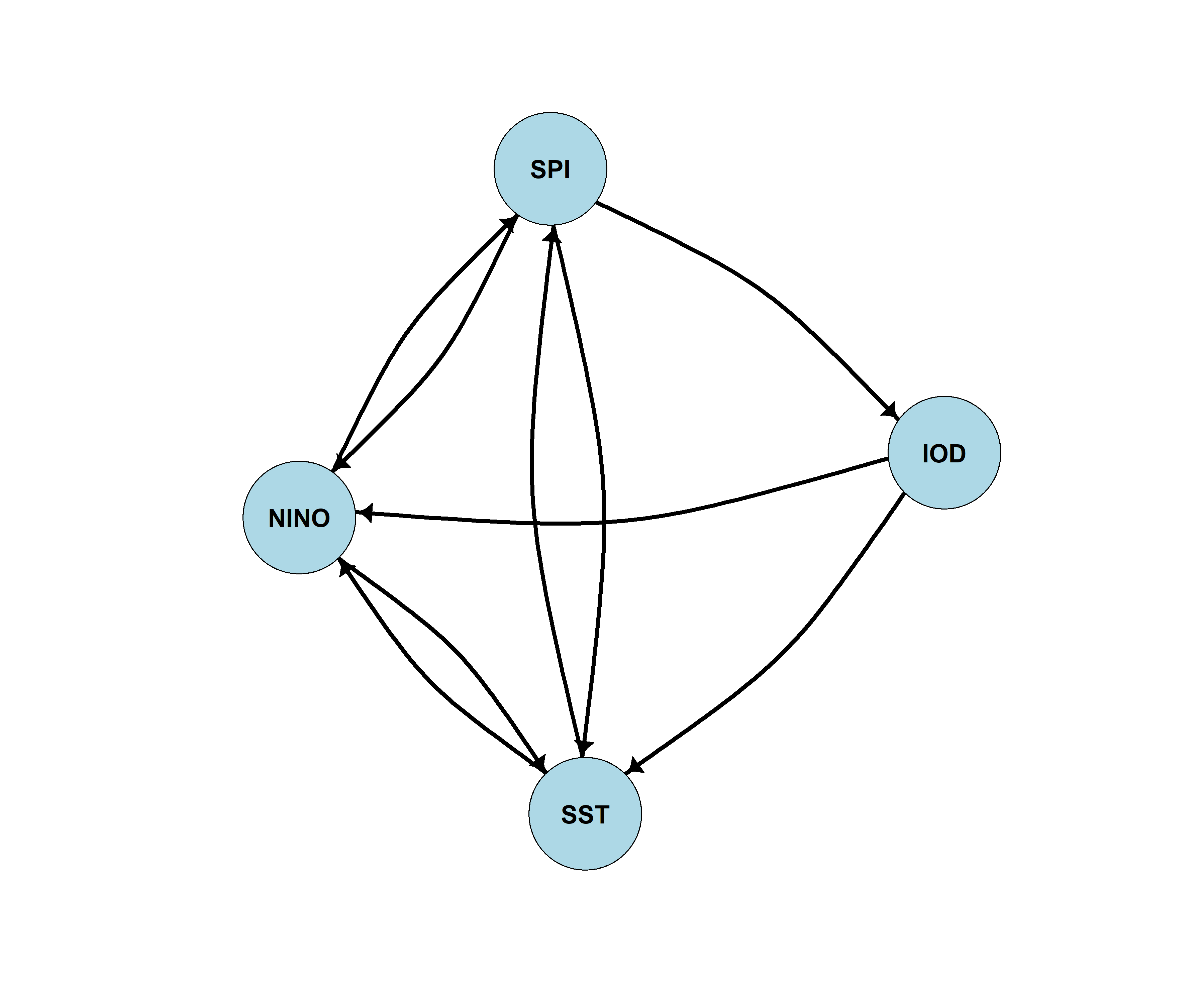}\\
 (a) & (b)\\
\end{tabular}
	\caption{Network of SST, NINO, IOD, and SPI based on Cross Correlation plots (\ref{fig_ccf_1999_2008},\ref{fig_ccf_2009_2018}) for (a) 1999 to 2008 and (b) 2009 to 2018.
	It is clear from the figure(a) that IOD has direct effect on Nino 3.4, SPI, and SST in decade 1998 t- 2008. But in decade 1999 to 2018 it does not have a direct effect on SPI. Whereas, Nino 3.4 and SST has direct effect in both the decades on SPI and vice-versa. Hence, IOD couple SPI indirectly. The  cross-correlation coefficient is significant enough to tell that all these climate variables have significant effect on each other.}
	\label{fig_network}
\end{figure}

To investigate the dynamics, we considered the cross-correlation functions (CCF) among the time-series of SPI, SST, IOD and NINO 3.4 variables see Figure~(\ref{fig_ccf_1999_2008}, \ref{fig_ccf_2009_2018}). The maximum lag is considered 120 months for this analysis. Figure (\ref{fig_ccf_1999_2008}) represents the CCF over 1999-2008, which shows that all the climate variables influence each other. Hence it creates a dense network, which we present in Figure (\ref{fig_network}a). Figure (\ref{fig_ccf_2009_2018}) presents the CCF over the period 2009-2018, where Figure~(\ref{fig_ccf_2009_2018}a) shows that SPI and SST influence each other. Figure~(\ref{fig_ccf_2009_2018}b) presents the CCF for SPI and IOD; it shows that the IOD does not couple SPI directly. Similarly, from Figure~(\ref{fig_ccf_2009_2018}c), we see that NINO 3.4 couples SPI and vice-versa. Figure~(\ref{fig_ccf_2009_2018}d) and (\ref{fig_ccf_2009_2018}e) show that IOD couple SST and NINO 3.4, but the reverse is not found 
Figure~(\ref{fig_ccf_2009_2018}f), we see that NINO 3.4 and SST couple each other. We present this analysis for 2009-2018 as a network diagram in Figure (\ref{fig_network}b).

\begin{figure}
\centering
\begin{tabular}{c|c}
 \includegraphics[width=.45\linewidth]{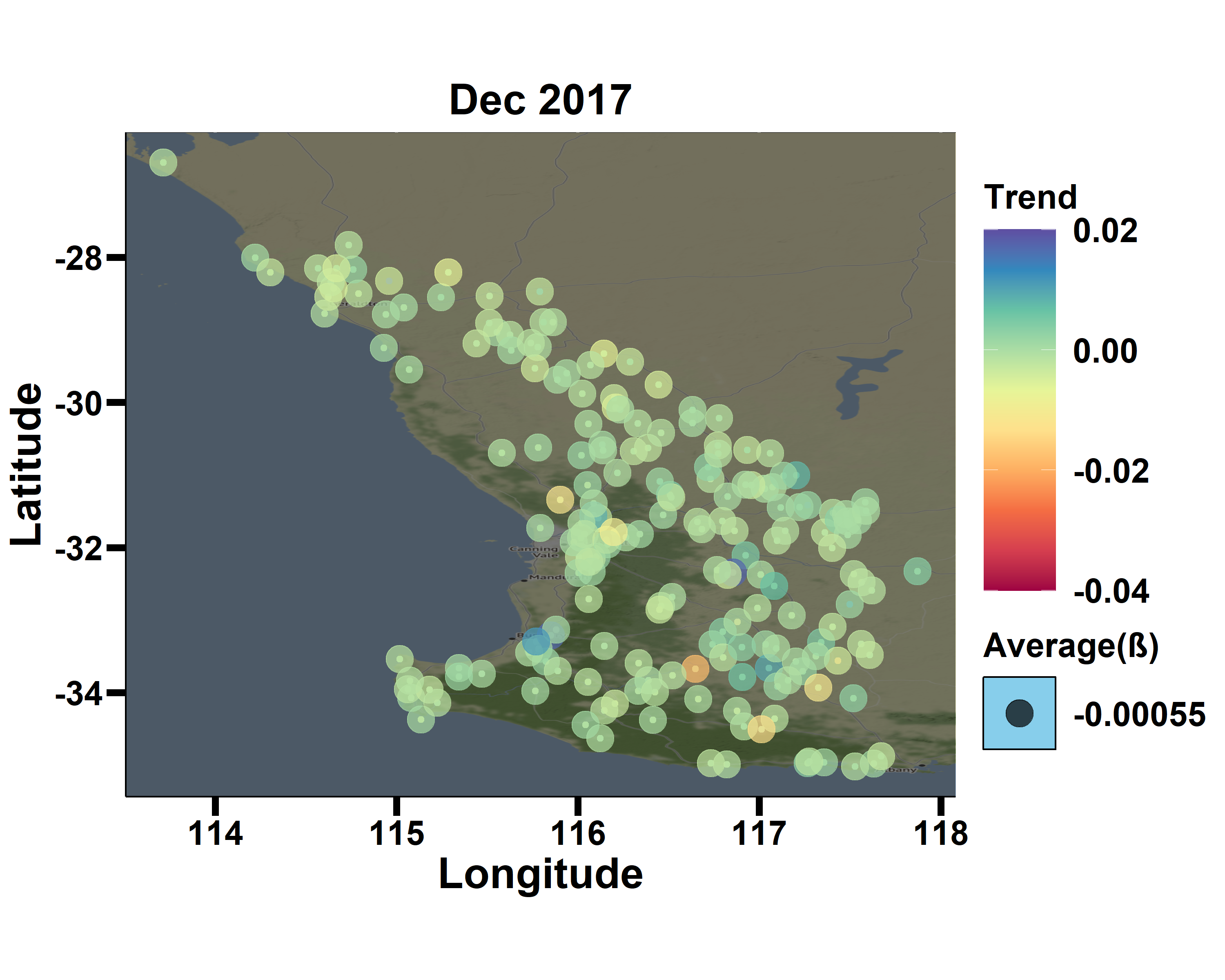}
 \includegraphics[width=.45\linewidth]{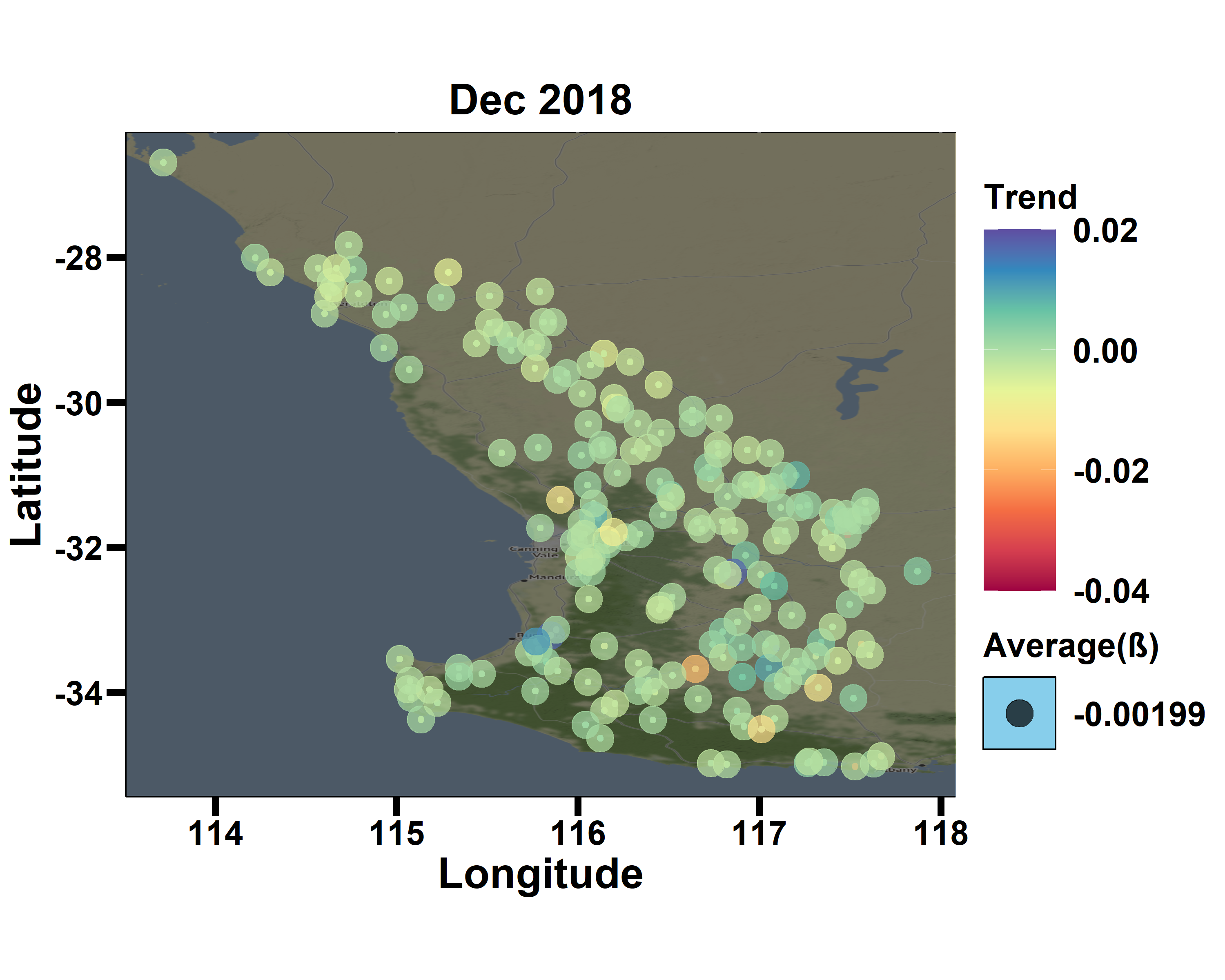}
\end{tabular}
    \caption{The trend ($\beta_1$) coefficient of SPI for Dec 2017 and Dec 2018 for 194 locations in the south-west Australia.}
 \label{fig_Trend_analysis}
\end{figure}

We developed a hybrid model that captures SPI's both long-term and short-term memory. The model identifies the long-term memory using Fourier harmonics. The Granger causal model captures the short-term memory and causality among SPI and other variables (NINO 3.4, IOD, and SST ). We also adjust the method by correcting spatial correlation among monitoring locations through the Gaussian process model. From the proposed model we  explored the trend in SPI to check if the study-area is becoming more drought-prone or wet prone? 
Equation (\ref{eqn_full_model}) estimates trends ($\beta_1$) for each year based on the past data. For example, Figure (\ref{fig_Trend_analysis}) presents the $\beta_1$ coefficients for Dec 2017 and Dec 2018 of all 194 locations. A location is more likely to see a dry weather if $\beta_1 < 0$, whereas, if $\beta_1 >0 $ then the location is more likely to see a wet condition. We observe that for both years, some locations show a mild increasing trend in the drought, especially in the inland. We also see that several locations near or close to the coastal area have $\beta_1 \approx 0$, indicating a non significant trend for Dec 2017 and 2018. 

\begin{figure}
    \centering

    \subfigure[]{\includegraphics[width=0.5\textwidth]{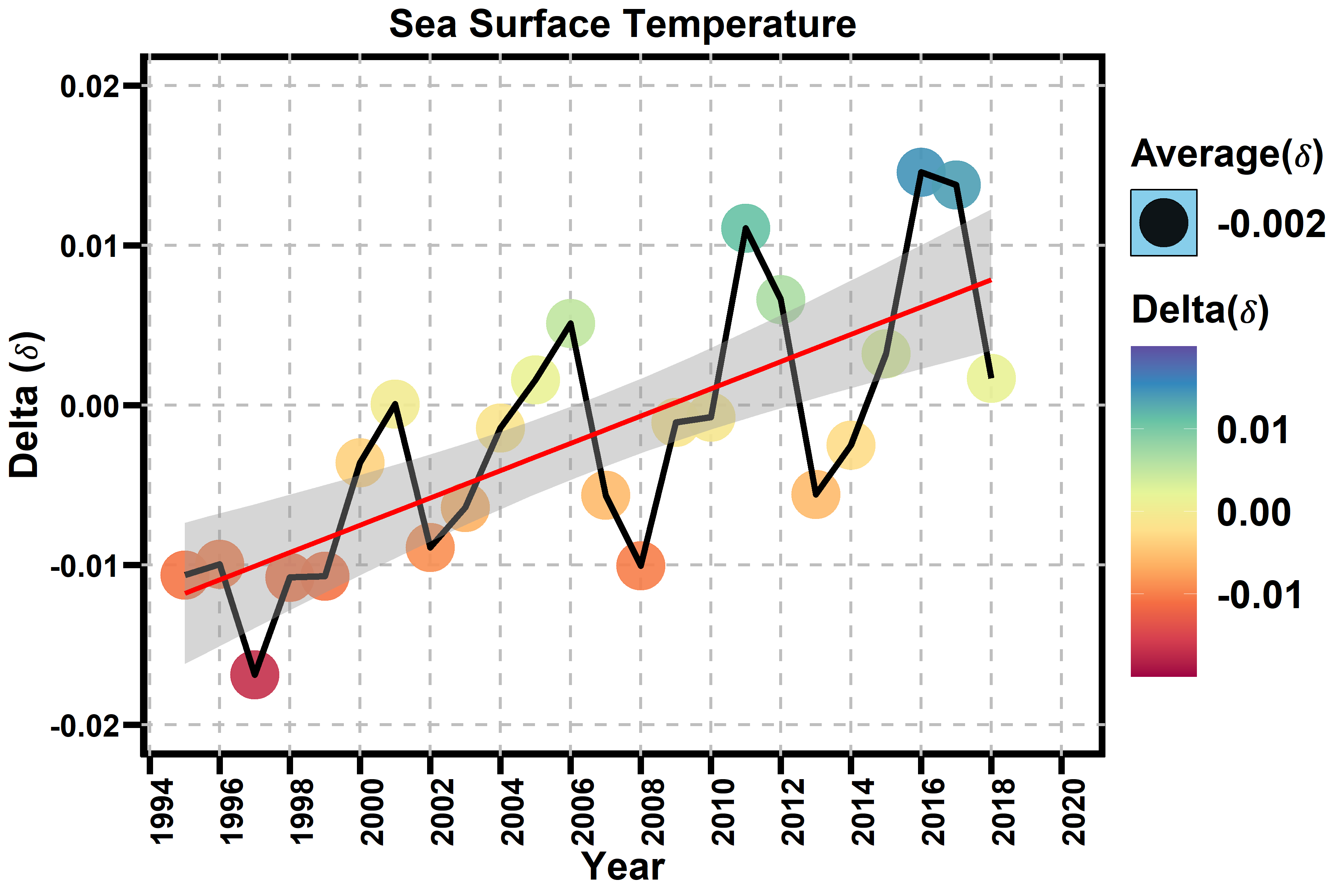}}
    \subfigure[]{\includegraphics[width=0.5\textwidth]{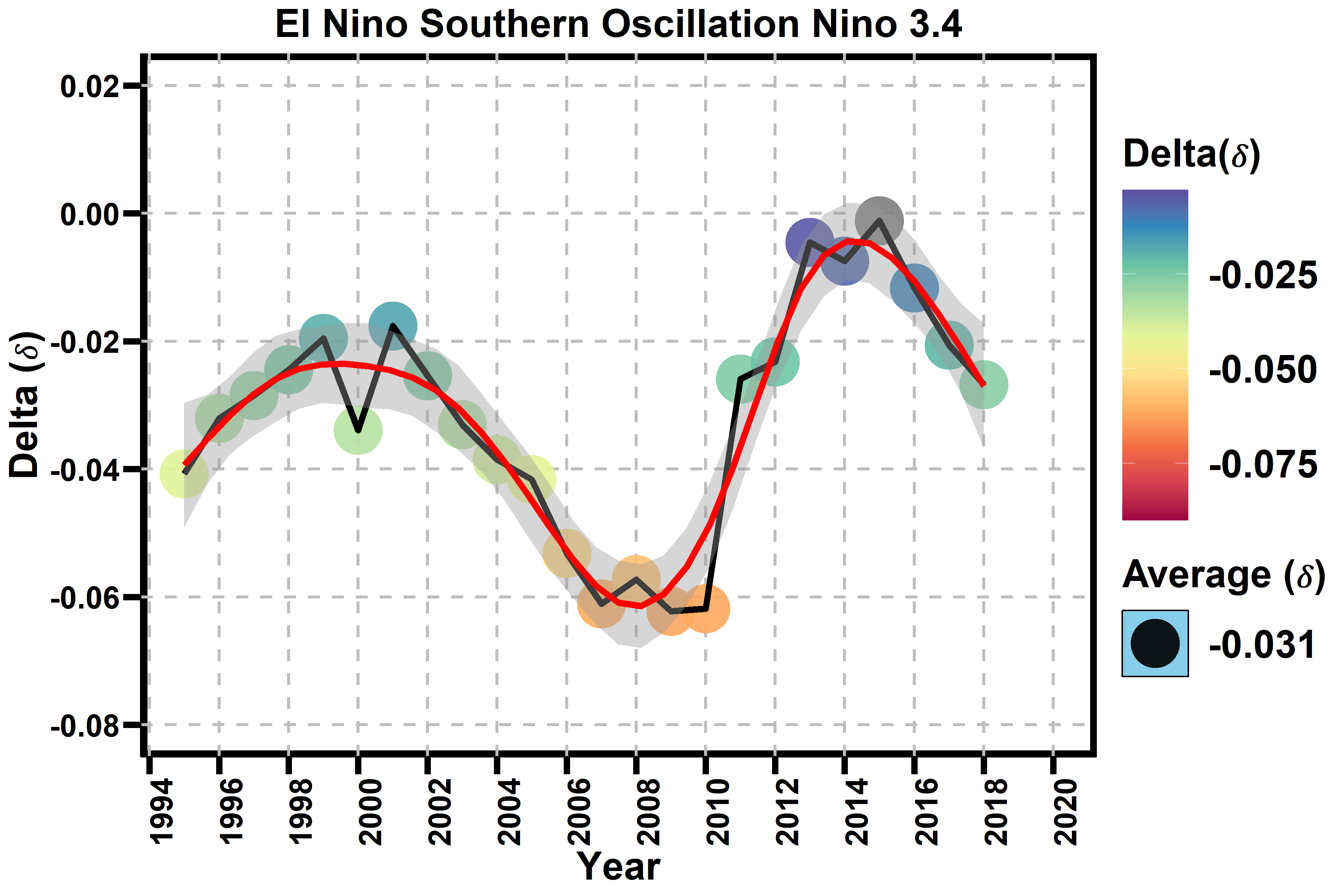}}
    \subfigure[]{\includegraphics[width=0.5\textwidth]{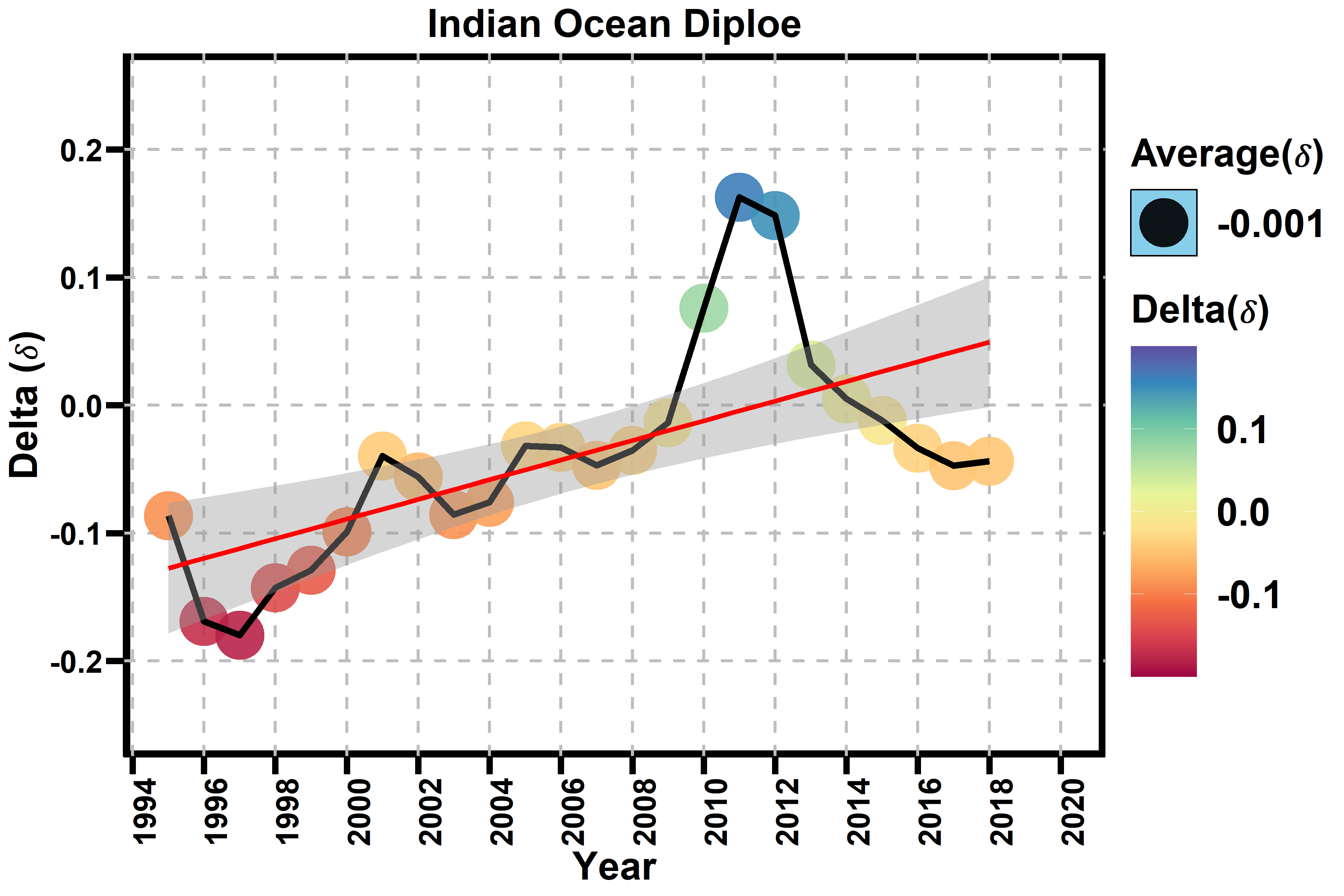}}
  
    \caption{ Plots for the average regression coefficient value ($\delta$) corresponding to NINO 3.4, SST, and IOD over the period 1995 to 2018. The red smooth curve is the three years moving average of $\delta$. It shows that NINO 3.4 always has a significant negative effect over SPI. On the other hand, until 2004, SST was always negatively correlated with SPI. During 2005 to 2104, the $\delta$ of SST swings between negative to a positive correlation. IOD was negatively correlated to SPI till the period of 2009 and in between 2009 to 2013 it was in positive range and after that it is significantly negatively correlated to SPI.}
    
     \label{fig_ts_beta_sst_nino}
    \end{figure}

Based on historical data \cite{IOD}, analysis presented in Figure (\ref{fig_network},\ref{fig_ts_beta_sst_nino}) showed that IOD and SPI were negatively correlated until 2008. Figure (\ref{ts_nino_sst_iod}b) shows a mild increasing trend in IOD. Increasing IOD results in decreasing SPI i.e., south-west Australia have faced more drought conditions around 2008.
After 2008, Figure (\ref{fig_ts_beta_sst_nino}c) indicates sharp increase in the regression coefficients of the IOD - in the positive region. It means that in the last decade (2009-2018), we have observed a reversal of relationship between IOD and SPI. If the trend in Figure(\ref{fig_ts_beta_sst_nino}c) continues in the positive direction, then we expect to see a more positive values of SPI, which means less dry and more wet conditions in south-west Australia. 
Similarly, from Figure (\ref{fig_ts_beta_sst_nino}b) we see that beta values of Nino 3.4 are all negative for all years, i.e., it has a negative relationship with SPI. That means, SPI values will be in positive range, which again means there will be more wet conditions in south-west Australia. 
Figure (\ref{fig_ts_beta_sst_nino}a) tells that SST and SPI are correlated to each other. As SST has increasing trend, SPI values will be in positive range, which means there will be more wet conditions in south-west Australia.

\begin{table}[p]
\begin{center}
\setlength{\arrayrulewidth}{1mm}
\setlength{\tabcolsep}{18pt}
\renewcommand{\arraystretch}{1.5}
\begin{tabular}{p{4cm}p{1cm}p{1cm}p{1cm}p{1cm}}\hline
 \bf{Different Combinations} & \bf{Type I}  & \bf{Type II}  &  \bf{Type III}& \bf{Type IV} \\  \hline
\textbf{Full Model without Spatial Correction} & 1.26 & 0.49 & 0.51& 0.38 \\ 
\textbf{Full Model with Spatial Correction} & 0.96 & 0.45 & 0.46 & 0.37 \\ 
\textbf{LASSO Selected Model with Spatial Correction} & 0.77 & 0.35 & 0.34 & 0.34 \\ 
 \hline
\end{tabular}

\caption{The out of sample Root Mean Square Error (RMSE) values for all four types of the model for from December 2010 to November 2018 (8 years). Here,  SPI is a target variable. The brief description of all types of the model are as follows: Type I: It captures the long term memory only using Fourier Series methods, Type II: Nino 3.4 and IOD are the covariates, Type III: lag values of Nino 3.4 and IOD are covariates, Type IV: Nino, IOD, and SST are covariates. We see that Type III and Type IV considering LASSO and GP correction are giving 0.34 the lowest (best) RMSE values in estimating SPI. }

\label{tab:RMSE}
\end{center}
\end{table}

\begin{figure}[p]
\centering
\includegraphics[width=0.8\linewidth]{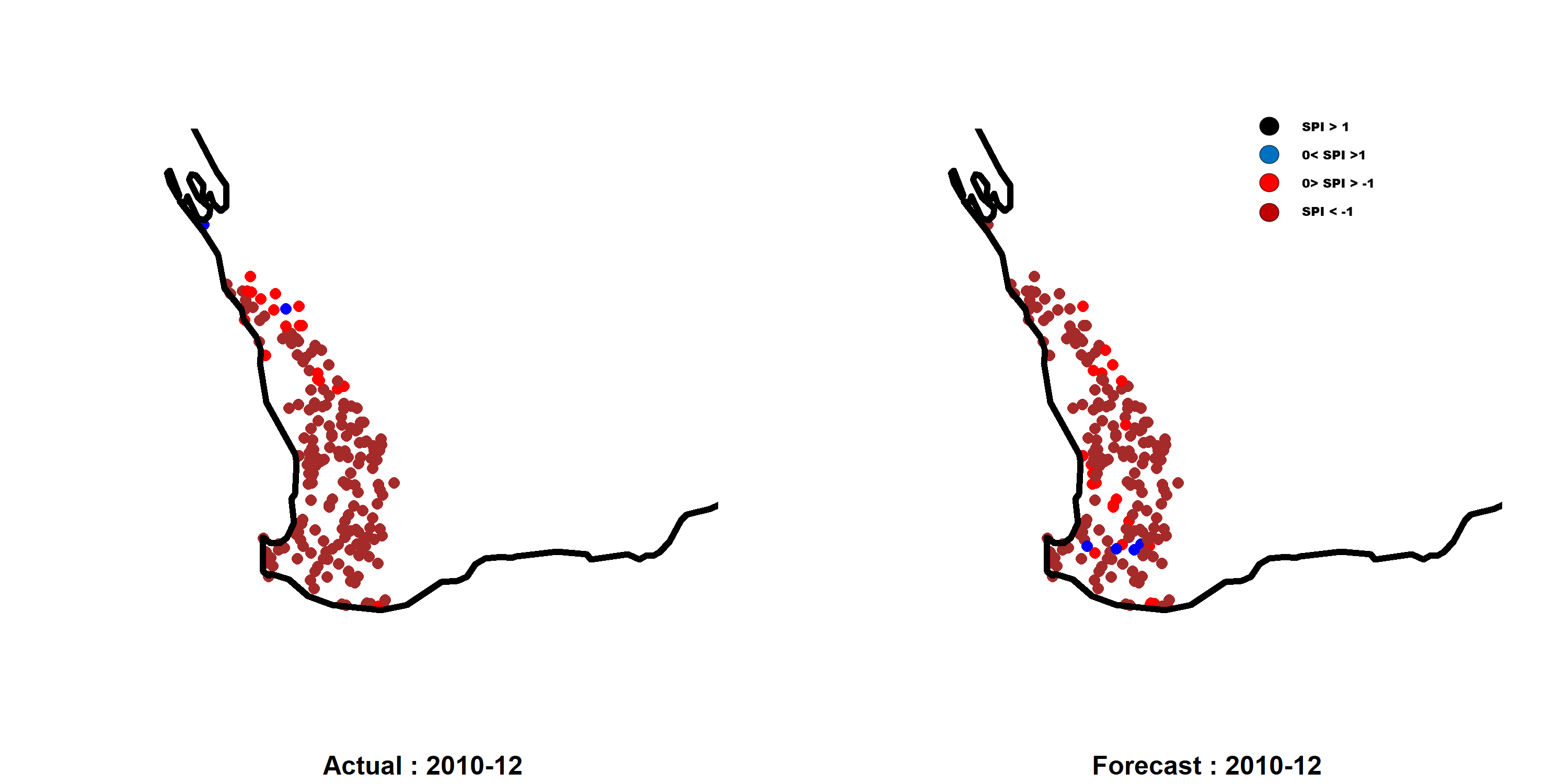}
\caption{estimate of SPI for Dec, 2010 using training data from June 1973 to November 2010. Defined following color scheme: (1) If SPI$>1$ : Black (2) If $1>$ SPI $>0$: \textcolor{blue}{Blue}; (3) If $-1<$ SPI $<0$: \textcolor{red}{Red}, (4) If SPI$<-1$: \textcolor{brown}{Brown}}
\label{fig_dec_2010}
\end{figure}

\section*{Methodology}\label{sec:methods}

To understand the complex dynamics of the climate variables NINO 3.4, IOD and SST with the SPI, we resort to different statistical machine-learning models. The Granger causal model \cite{Granger1969} is useful to determine whether one-time series helps to estimate another. It is also helpful in capturing the short-term memory of the system. We developed it to test the causal relationship among SPI, SST, NINO 3.4, and IOD. We used the Fourier series model to investigate the SPI's long-term memory for all rainfall monitoring locations. The Fourier series method used here is a special case of the functional data analysis technique to capture the long-term memory \cite{Das2018}. For each location, we develop a hybrid statistical model, which captures the long-term behaviour with the Fourier series method and the Granger causal model describes the short-term behaviour. We also correct the estimates for spatial correlation among the locations using the Gaussian Process (GP) model. 
We used the following algorithms to identify the long memory period of each chain.

\begin{figure}[p]
\centering
    \includegraphics[width=.5\linewidth]{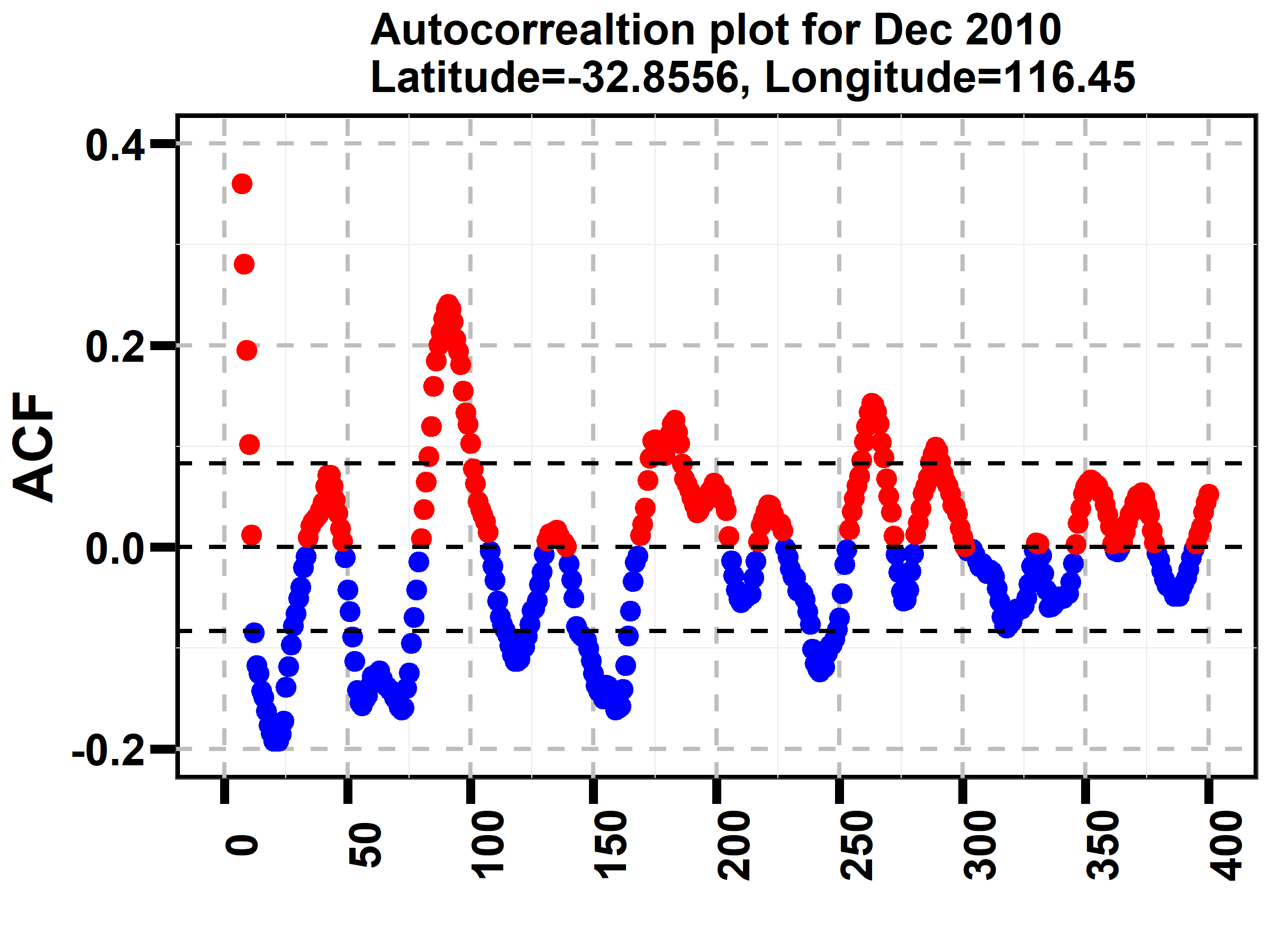}
\caption{The autocorrelation plot with a maximum lag of 400 months for a rainfall gauged location at the longitude 116.45 and latitude -32.86. We used 450 months of data, from June 1973  to Nov 2010, to create the autocorrelation function. There are four significant periods of 91, 183, 263, and 289 months. The first three periods indicate that the current value of SPI has a significant positive correlation with a past SPI value with a periodicity of about 7.5 years.}
\label{fig_acf_9575}
\end{figure}

\begin{enumerate}
    \item[(i)] Calculate autocorrelation function on training time series data.
     \item[(ii)] Identify period $P_1, P_2,\cdots, P_{m_s}$, with autocorrelation $\rho_m>s$, where\\
	 $s=\frac{1}{M}\sum_{m=1}^M|\rho_m-\hat{\rho}|$; $\rho_m$ is the $m^{th}-$lag autocorrelation; $\hat{\rho}$ is median of all autocorrelation; $M$ is the maximum lag considered in the study.
\end{enumerate}
Figure (\ref{fig_acf_9575}) shows the autocorrelation plot for a location (coordinate:  longitude 116.45 and latitude -32.85), which has a maximum lag of 400 months. We used 450 months of data, from June 1973 to Nov 2010, to create the autocorrelation function. By implementing the algorithm, we identify four periods: 91, 183, 263, and 289 months. The first three periods indicate that the current value of SPI has a significant positive correlation with a past SPI value with a periodicity of about 7.5 years 
Detail investigation indicates that for all 194 locations, the period number varies where average Value is around 148 months (approx 12 years) with standard deviation 103.15 months. Figure (\ref{fig_acf_9575}) is a representative figure of one out of 194 locations in the study area. We repeated this process for all 194 locations and identified the long memory periodicity of each location.


\begin{figure}[p]
    \centering
    \includegraphics[width=.5\linewidth]{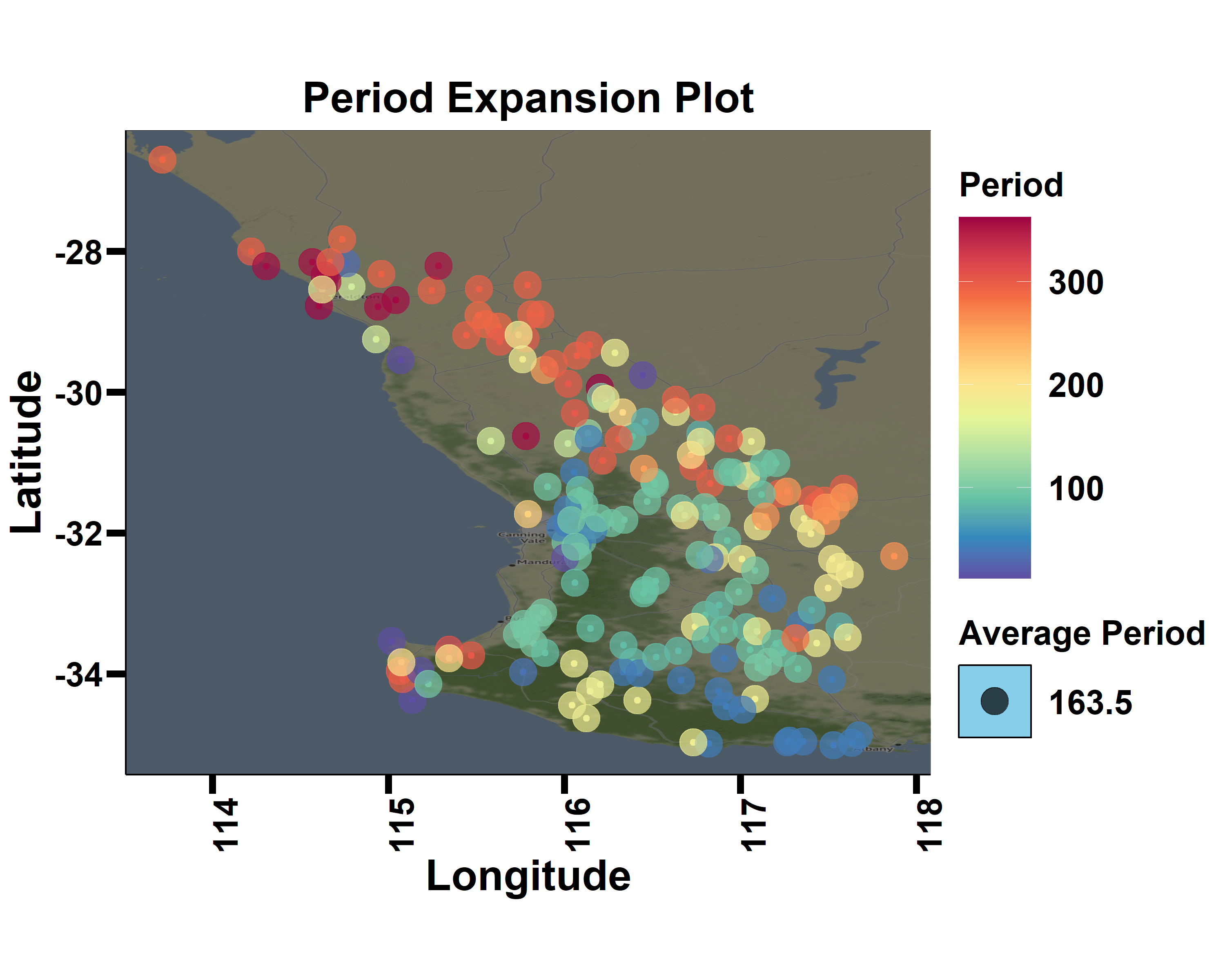}
\caption{The plot presents the long-memory period for each of the 194 locations. There is a strong spatial correlation between all locations' long-memory periods. The colour bar shows the length of the period in the number of months. }

    \label{fig_Period1_194}
\end{figure}

Figure (\ref{fig_Period1_194}) shows the long-memory period for all 194 locations on the western Australia map. The long memory period ranges from 60 months to 360 months, and clearly, we see a strong spatial correlation. The identification of the period would help us in analyzing the seasonality. For each location, we consider 450 months of data, starting from June 1973 to Nov 2010. The rest of the data from December 2010 to 2018 were used to test the generalization of the analysis. We propose the following hybrid statistical machine learning model for location $s = 1,2,\cdots, S(=194),$
    \begin{eqnarray}
       \label{eqn_full_model} M_s:~~~y(t)=\beta_0+ \beta_1 t+\alpha(t) + \eta(t)  +
        W(s)+\epsilon(t),
    \end{eqnarray}
where $\beta_0$ is intercept, $\beta_1$ is the coefficient of trend, $\alpha(t)$ models the short-term memory of the process, $\eta(t)$ models the long term memory of process. The $W(s)$ is the spatial or geographical effect at location $s$, where $W(s)\sim GP(0,\Sigma(s,s'))$, i.e., $W(s)$ follows Gaussian Process with mean zero and covariance function as: $\Sigma(s,s')=\tau^2\exp\{-\rho|s-s'|^2\}$, where we consider an exponential covariance matrix $\exp\{-\rho|s-s'|^2\}$. The $\epsilon(t)$ is the white noise with $\mathbb{E}(\epsilon)=0$ and $\mathbb{V}ar(\epsilon)=\sigma^2$, $M_s$ denote the model for location $s$.
We consider four different types of the model, as follows: 

\begin{itemize}
\item \textbf{Type I}: Long term memory model
    \begin{eqnarray}
    \label{eqn_model1}
      \eta(t)=\sum_{j=1}^{m_s}&\bigg\{&\sum_{i=1}^K \beta_{ji}\sin(i*\omega_j*t)  +\sum_{i=1}^{K}\gamma_{ji}\cos(i*\omega_j*t)\bigg\},
    \end{eqnarray}
where $\omega_j=\frac{2\pi}{P_j},~~ j =1,2,\cdots,m_s$, $P_j$ is estimated via the algorithm explained earlier, and $m_s$ denote the number of periods for $s^{th}$ location. 
For the short term memory, i.e. $\alpha(t)$, we considered two different approaches considering Granger causal model: 

\item \textbf{Type II}: Short term memory model with Nino3.4 and IOD as estimators.
Here, $y(t-K)$ denotes as $k^{th}$ lag SPI and $ X(t)$ as a estimator's time-series such as NINO 3.4 and IOD.
\begin{equation}
\label{eqn_model2}
  \alpha(t) = \alpha_0 + \alpha_1 y(t-k) + \delta X(t),  
\end{equation}

\item \textbf{Type III}: Short term memory model with the lag effect of Nino3.4 and IOD.\\
In this model we take lag effect for $X(t)$ (Nino and IOD) such that 
\begin{equation}
\label{eqn_model3}
 \alpha(t) = \alpha_0 + \alpha_1 y(t-k) + \delta X(t-k),   
\end{equation}

where $y(t-K)$ is the $k^{th}$-lag SPI, $X(t-k)$ is the $k^{th}$-lag estimator's time-series.

\item \textbf{Type IV}: Short term memory model with Nino3.4, IOD and SST. \\
In this model we take $ X(t)$ as a estimator's time-series such as NINO 3.4, SST and IOD.
\begin{equation}
\label{eqn_model3}
 \alpha(t) = \alpha_0 + \alpha_1 y(t-k) + \delta X(t-k),   
\end{equation}
\end{itemize}

We implemented the machine learning shrinkage technique the LASSO (least absolute shrinkage and selection operator) selection to find out the best harmonics in the Fourier model \cite{LASSO1996}. The LASSO technique only picks the harmonics that have a statistically significant effect in reducing the error. We fit the model for location $s$, with data from $\{(t-w),\cdots,(t-1)\}$, where $w=450$ months.
For estimating $y(t)$ at the location $s$, we define the estimated value as $\hat{y}_s(t)$ for $t^{th}$ month, and apply spatial correction with GP model as
    \begin{equation*}
    \bar{y}(t)=\Sigma(s,s')[\Sigma(s,s')+\tau^2\mathbf{I}]^{-1}\hat{y}(t),
    \end{equation*}
where $\Sigma(s,s')=exp\{-\rho|s-s|^2\}$ as defined in Equation (\ref{eqn_full_model}). 

The LASSO approach for the Fourier model with spatial correction helps improve the model's performance. The SST contributes significantly to improving the accuracy of out-of-sample SPI. For the Fourier model without spatial correction, we see that without a short-term memory, the value of Root Mean Squared Error (RMSE) is 1.26 (Full Model without Spatial correction, Type I). However, when we considered short-term memory, the RMSE was reduced to 0.49 for Type II and 0.51 for Type III. By considering SST as covariate in the Type III, we see that the out-sample RMSE dropped to 0.38. It means that SST has a significant effect in estimating SPI. When we considered the spatial GP correction without LASSO, the RMSE value dropped from 1.26 to 0.96 (Type I), and it decreased from 0.38 to 0.37 when we consider SST as a covariate (Type 1V). Similarly, we see that with GPs spatial correction and Lasso selection for optimal Fourier harmonics, the RMSE value is reduced further to 0.77 (Type I) and the out-sample RMSE is the lowest level of 0.34 for Type III and Type IV. Our analysis shows that the LASSO selection process with spatial correction model can improve the accuracy of the proposed model by many folds as RMSE value is lowest (0.34) in this case.

\section*{Discussion and Summary}
\label{sec:5}
We developed the hybrid model and analyzed the relationship of the SPI to SST, NINO 3.4 and IOD. We used the machine learning algorithm including LASSO to select the significant Fourier harmonics to model the long memory. To capture the short term memory of SPI we used lagged estimators like IOD, SST and NINO 3.4 in Granger causal model. We also applied the GP spatial correction in estimating SPI. Our study observed that SPI, SST, NINO 3.4 and IOD  have long memory as Hurst exponent values are estimated above 0.5, see Table (\ref{tab_Hurst_exp}). 
We have performed a number of validations of the proposed model based on out of sample test data. In validation step, we set aside the month that we want to validate. For example, Figure (\ref{fig_dec_2010}) represents the actual and estimated value in Dec, 2010 on Australia map. Here, we take training data from June, 1973 to Nov, 2010, i.e., 450 months, and we apply the proposed model (\ref{eqn_full_model}) to estimate the SPI value for Dec, 2010. In Figure(\ref{fig_dec_2010}), we presented the out of the sample estimates and actual SPI. The visual inspection also indicates that the proposed model(\ref{eqn_full_model}) estimated well for Dec 2010. However, this validation is just for one particular month. To evaluate the performance of the proposed model, we repeat this estimation process from December 2010 to November 2018 (for eight years) and calculate the out-of-sample RMSE of the estimates for 194 locations. The results for the median RMSE are given in Table (\ref{tab:RMSE}). 

 Based on the best performing model, we observed an increasing trend in SST attribute change in the characteristics of the SPI's distribution. Figure (\ref{fig_ts_beta_sst_nino}) shows the average regression coefficient value ($\delta$) corresponding to SST, IOD and Nino 3.4 over the period from 1995 to 2018. It indicates that Nino 3.4 always has a significant negative effect over SPI. IOD was negatively correlated to SPI till the period of 2008 and in between 2009 to 2013 it was in positive range and after that it is significantly negatively correlated to SPI. However, until 2004, SST was always negatively correlated with SPI. During 2005 to 2014, the SST has some swings between negative to a positive correlation. Since 2014, we observe that the regression coefficient ($\delta$) corresponding to SST is always positive. Overall, we observe an upward trend in the $\delta$ corresponding to SST in Figure (\ref{fig_ts_beta_sst_nino}). It means, as we know SST itself has an upward trend, the upward positive trend of $\delta$ indicates that SPI is positively correlated with SST in recent years. This implies a more wet season in the study area of Western Australia.
 Though 12-month moving average of SPI has a negative trend towards drought, see Figure (\ref{fig_ts_SPI}b), but the complex dynamics of SPI with other climate variables indicates more wet season for western Australia.
 




\bibliography{bibliography}

\section*{Acknowledgements}

A.Y. is grateful for the fellowship from JNU and hospitality at CMI funded by AlgoLabs. S.D. acknowledges the partial financial support from Infosys Foundation, TATA Trust, and Bill \& Melinda Gates Foundation's grant to CMI.

\section*{Author contributions statement}

 A.C., S.B., S.D., and A.Y. conceived the research. S.D. and A.Y. developed the methods, analysed the results and prepared the figures. All authors discussed the results, contributed to the writing of the manuscript, and reviewed the manuscript. 

\section*{Ethics declarations}
\textbf{Competing interests} : The authors declare no competing interests. 


\end{document}